\documentclass[galaxies,article,accept,moreauthors,latex,10pt,a4paper]{mdpi}
\firstpage{1}
\articlenumber{x}
\doinum{10.3390/------}
\pubvolume{5}
\pubyear{2017}
\copyrightyear{2017}
\history{Received: 9 October 2017; Accepted: 28 November 2017; Published: date }

\usepackage{verbatim} 
\usepackage{amssymb}

\Title{Multi-Frequency VLBA Polarimetry and the Twin-Jet Quasar 0850$+$581 $^\dagger$}

\Author{Evgeniya Kravchenko $^{1,}$* and Yuri Y. Kovalev $^{1,2,3}$}
\AuthorNames{Evgeniya Kravchenko and Yuri Y. Kovalev}

\address{%
$^{1}$ \quad Lebedev Physical Institute, Astro Space Center, Profsoyuznaya 84/32, 117997 Moscow, Russia; yyk@asc.rssi.ru\\
$^{2}$ \quad Moscow Institute of Physics and Technology, Institutsky per., 9, 141700~Dolgoprudny, Moscow~Region,~Russia\\
$^{3}$ \quad Max-Planck-Institut f\"ur Radioastronomie, Auf dem H\"ugel 69, 53121 Bonn, Germany}

\corres{Correspondence: evk@asc.rssi.ru}

\abstract{
We present the first multi-frequency VLBA
 study of the quasar 0850$+$581 which appears to have a two-sided relativistic jet.
Apparent velocity in the approaching jet changes from 3.4c to 7c with the separation from the core.
The jet-to-counter-jet ratio of about 5 and apparent superluminal velocities suggest that the observing angle of the inner jet is $\leq17^\circ$.
It is likely that this orientation significantly changes downstream due to an interaction of the jet with the surrounding medium; signs of this are seen in polarization.
A dense inhomogeneous Faraday screen is detected in the innermost regions of this quasar.
We suggest that there is a presence of ionized gas in its nucleus, which might be responsible for the free-free absorption of the synchrotron emission in the jet and counter-jet at frequencies below 8.4~GHz.
The experiment makes use of slowly varying instrumental polarisation factors (polarization leakage or D-terms) in time.
We report application of the ``D-term connection'' technique for the calibration of an absolute orientation of electric vector position angle (EVPA) observed by VLBA at 4.6, 5.0, 8.1, 8.4, 15.4, 22.3, and 43.3~GHz bands during the 2007--2011.
}


\conference{Polarised Emission from Astrophysical Jets}

\begin{document}

\section{Introduction}
Active galactic nuclei (AGN) are among the most interesting space objects: they host supermassive black holes, creating natural laboratories with extreme physical conditions.
Good candidates to study this phenomenon are relativistic jets, ejected from the central regions of AGNs.
They~are produced by strong magnetic fields \citep{2001Sci...291...84M}, which give rise to synchrotron radiation, collimation, acceleration, plasma instabilities and other effects.
Polarimetric Very Long Baseline Interferometric (VLBI) observations deliver a wealth of information on them and play a crucial role in the study of AGN jets.


The calibration of polarimetric VLBI observations is a challenge.
These require (a) a good parallactic angle coverage for suitable calibrator to determine instrumental polarization and (b)~comparison of its electric vector position angle (EVPA) orientationone defined within monitoring programs to calibrate absolute orientation of the polarization angle.
Roberts et al. \citep{roberts_etal94} that antenna-based right and left (R and L) circular polarization amplitudes and phases vary slowly with time at 5 and 8~GHz. %
They suggested to connect close-in-time leakage solutions to calibrate the absolute R-L phase difference, which makes it unnecessary to use appropriate calibrators or to compare EVPA with monitoring programs.
Later, \cite{1995AJ....110.2479L,gomez_etal02} supported their results for observations at 22 and 43~GHz and showed the efficiency of this technique in increasing of calibration accuracy for individual experiments.

Here, we present time behavior (Section~\ref{s:1}) of the D-term amplitudes and phases for the VLBA antennas from 4 to 43~GHz, obtained between 2008 and 2011.
We used small temporal variability of these values to better estimate the R-L phase offset.
This approach is justified for experiments with polarization calibrators, whose EVPA orientation changes quickly in time, while the closest available EVPA measures is a few weeks.
Moreover, missed IFs or antennas still may be calibrated.

This approach is used to calibrate observational data of the quasar 0850$+$581, located at $z=1.318$ and demonstrating unusual polarization properties.
Study of the object was motivated by the results of opacity effect measures in 277 radio sources by \cite{kovalev_etal08}.
The quasar showed the largest apparent frequency-dependent core shift in a sample between 2 and 8~GHz, which amounts to (1.35 $\pm$ 0.03)~mas.
For understanding of an underling physical processes and conditions in the source, we recorded our observations with the VLBA simultaneously at 4--43~GHz in full polarization mode in 2008.
The object has not been studied before in detail. Here, we present first results of its observations, Section~\ref{s:2}.

The angular scale of one mass corresponds to 8.45~pc, assuming cosmology with Hubble constant $H_\mathrm{0}$ = 71~km/s/Mpc, the vacuum energy density $\Omega_{\lambda}$ = 0.73 and matter density $\Omega_\mathrm{M}$ = 0.27.

\section{Temporal Behavior of D-Term Amplitudes and Phases}
\label{s:1}

The observations reported here represent a set of individual VLBA projects
made at 4.6, 5.0, 8.1, 8.4, 15.4, 22.3 and 43.3~GHz in full Stokes.
A summary is given in Table~\ref{t:1}. Data~includes 29 epochs,

\begin{table}[H]
\caption {The VLBA projects are used to connect D-term phases and amplitudes.}
\label{t:1} %
\centering
\scalebox{.85}[0.85]{\begin{tabular}{cccc}
\toprule
\textbf{Project Code}	& \textbf{Date}	& \textbf{Calibrator} & \textbf{Reference}\\
\midrule 
BK134  & 1 March 2007, 30 April 2007, 3 May 2007, 1 April 2007 & 4C~$+$39.25 & E.K. \cite{kravchenko_etal17}\\\midrule %
BK142  & 17 February 2008               & 0833$+$585  & E.K. (This paper,\\
       & 		            & 4C~$+$39.25 & E.K. et al. in prep.)\\\midrule
BK150  & 2 September 2008, 5 September 2008, 2 October 2008, 23 October 2008 & 3C454.3     & E.K. \\
       & 2 February 2009, 5 February 2009, 9 April 2009, 22 April 2009,&  OJ287  & \\
       & 14 May 2009 &             & \\
       & 1 June 2009, 20 June 2009 & 3C454.3     & \\\midrule
SE2087A& 28 August 2009, 25 October 2009, 5 December 2009, 26 January 2010 & OP~313      & M.Lisakov \cite{lisakov_etal17},\\
       &  &             &  Lisakov et al. in prep.\\\midrule
SE2087B& 22 September 2009, 22 October 2009, 3 December 2009, 18 January 2010, & 3C454.3     & T.Savolainen\\
       & 21 February 2010 &             & \\\midrule
SE2087E& 24 May 2010, 9 July 2010, 28 August 2010, 18 October 2010 & OJ~287      & E.K. \cite{kravchenko_etal16}\\
\bottomrule
\end{tabular}}
\end{table}

\noindent which cover the period from 2007 to 2011.
The observations were calibrated within AIPS \cite{bridle_greisen_94}, while polarization leakage parameters were estimated using the linear feed solution algorithm \citep{1995AJ....110.2479L}, implemented within the LPCAL task.
The bands at 4.6--8.4~GHz are split into 4$\times$8~MHz and 15.4--43.3~GHz bands---into 2$\times$16~MHz-wide frequency channels (IFs).
The instrumental polarization is determined separately for each IF.
The integrated EVPA orientations were referenced to the values, determined within the following programs:
\begin{itemize}[leftmargin=*,labelsep=5.5mm]
\item VLA Monitoring Program\footnote{\url{http://www.vla.nrao.edu/astro/calib/polar/}.} \cite{taylor_myers_00};
\item the University of Michigan Radio Astronomy Observatory monitoring program\footnote{\url{https://dept.astro.lsa.umich.edu/datasets/umrao.php}.};
\item 43~GHz VLBA data from the VLBA-BU Blazar Monitoring Program \footnote{\url{http://www.bu.edu/blazars/VLBAproject.html}.};
\item MOJAVE program \cite{lister_etal09}.
\end{itemize}

The approximate time separation between ours and these measures is a few weeks and in some cases reaches six months.
Because of this sparse time coverage, this work was initiated.

\begin{figure}[H]
\centering
\includegraphics[width=6 cm]{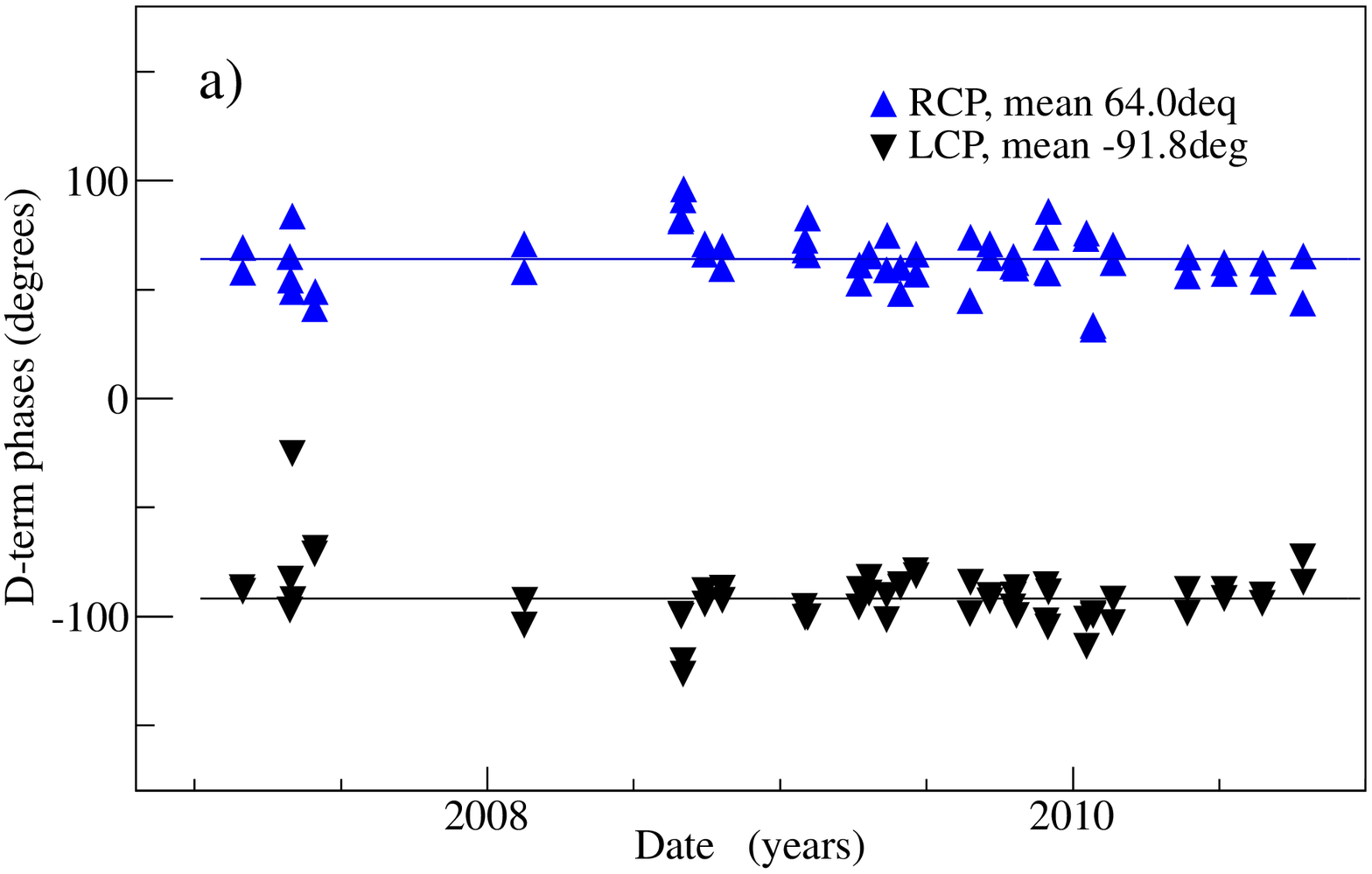}\quad
\includegraphics[width=6 cm]{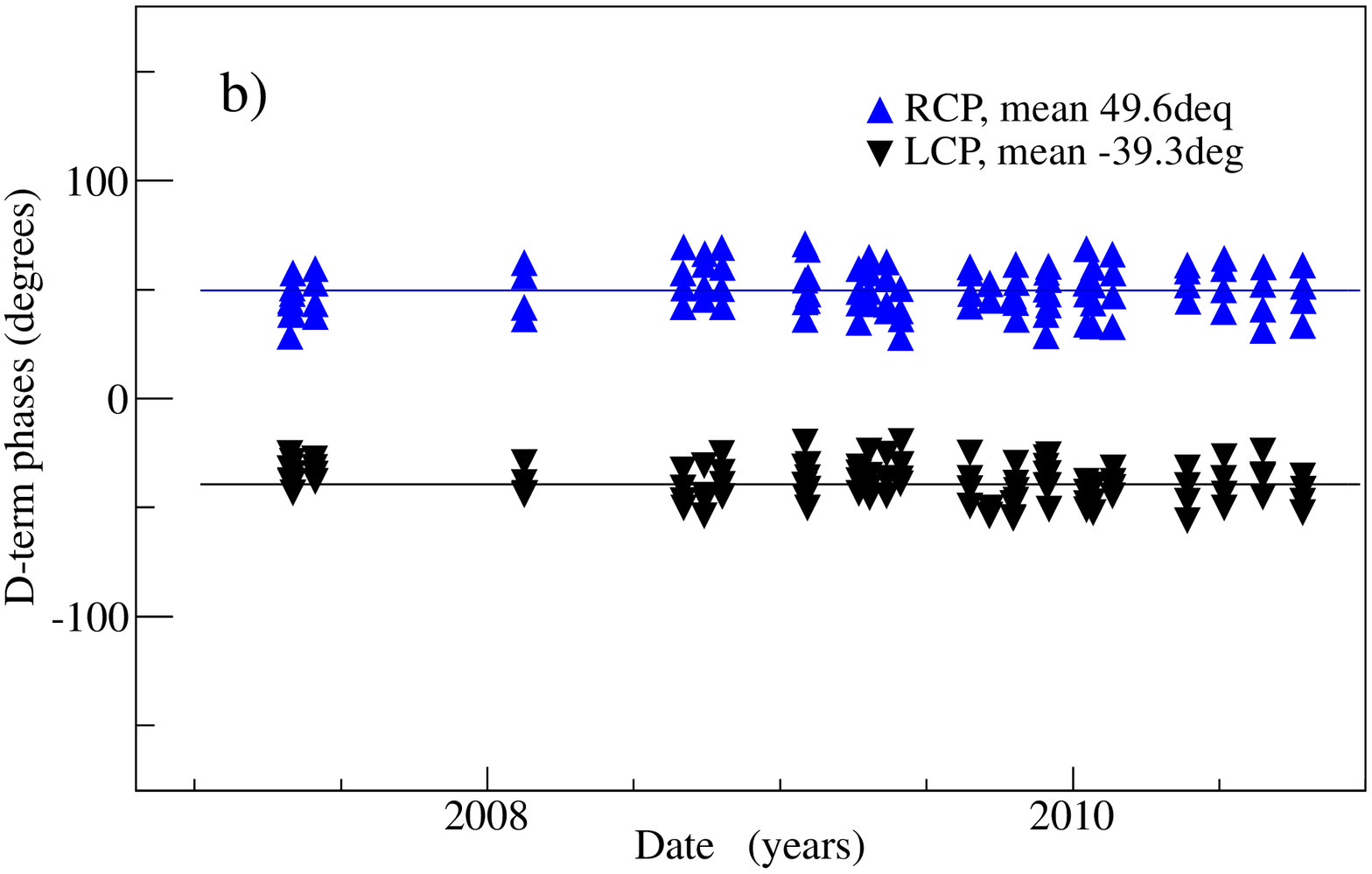}\\
\includegraphics[width=6 cm]{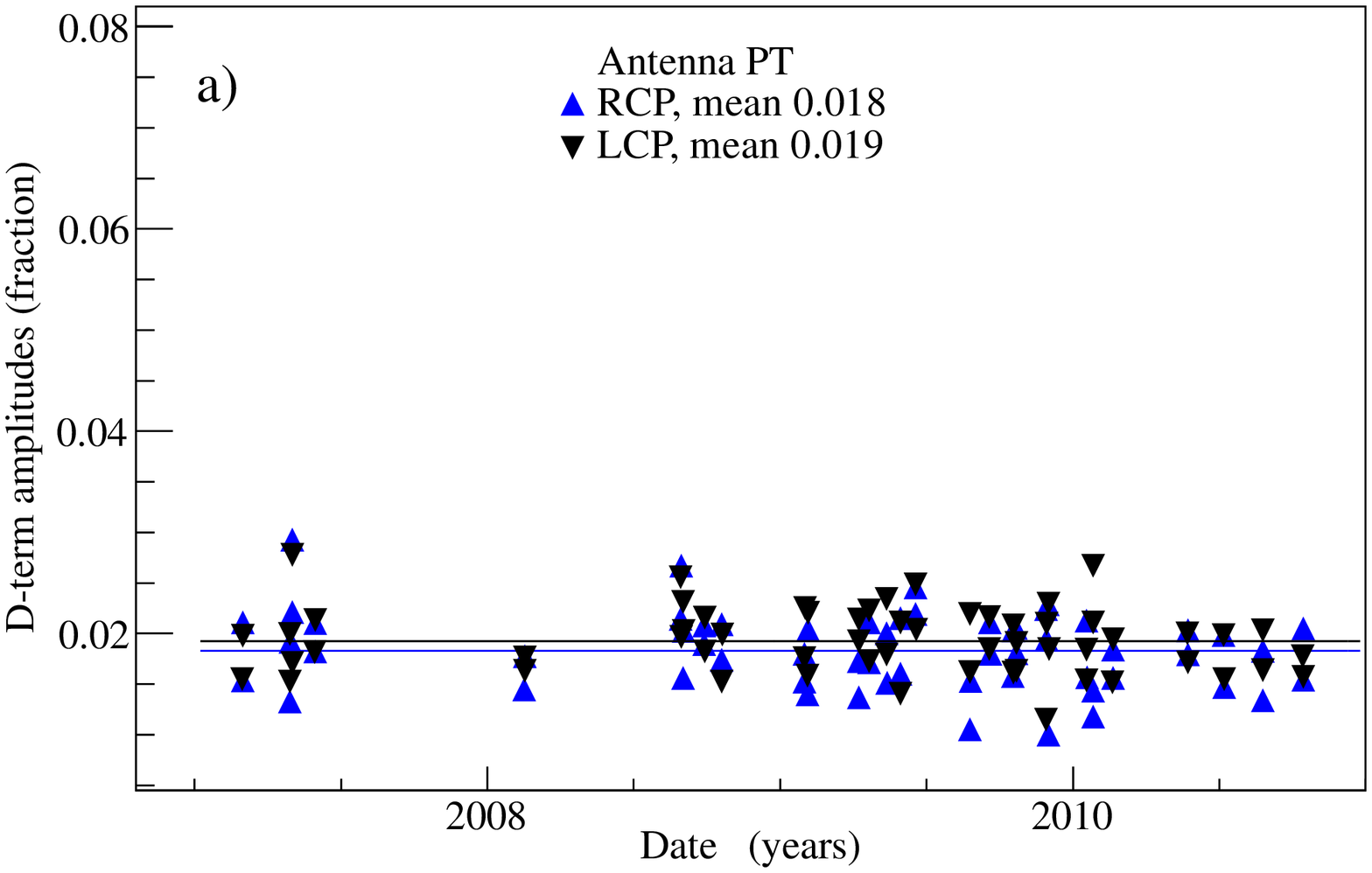}\quad
\includegraphics[width=6 cm]{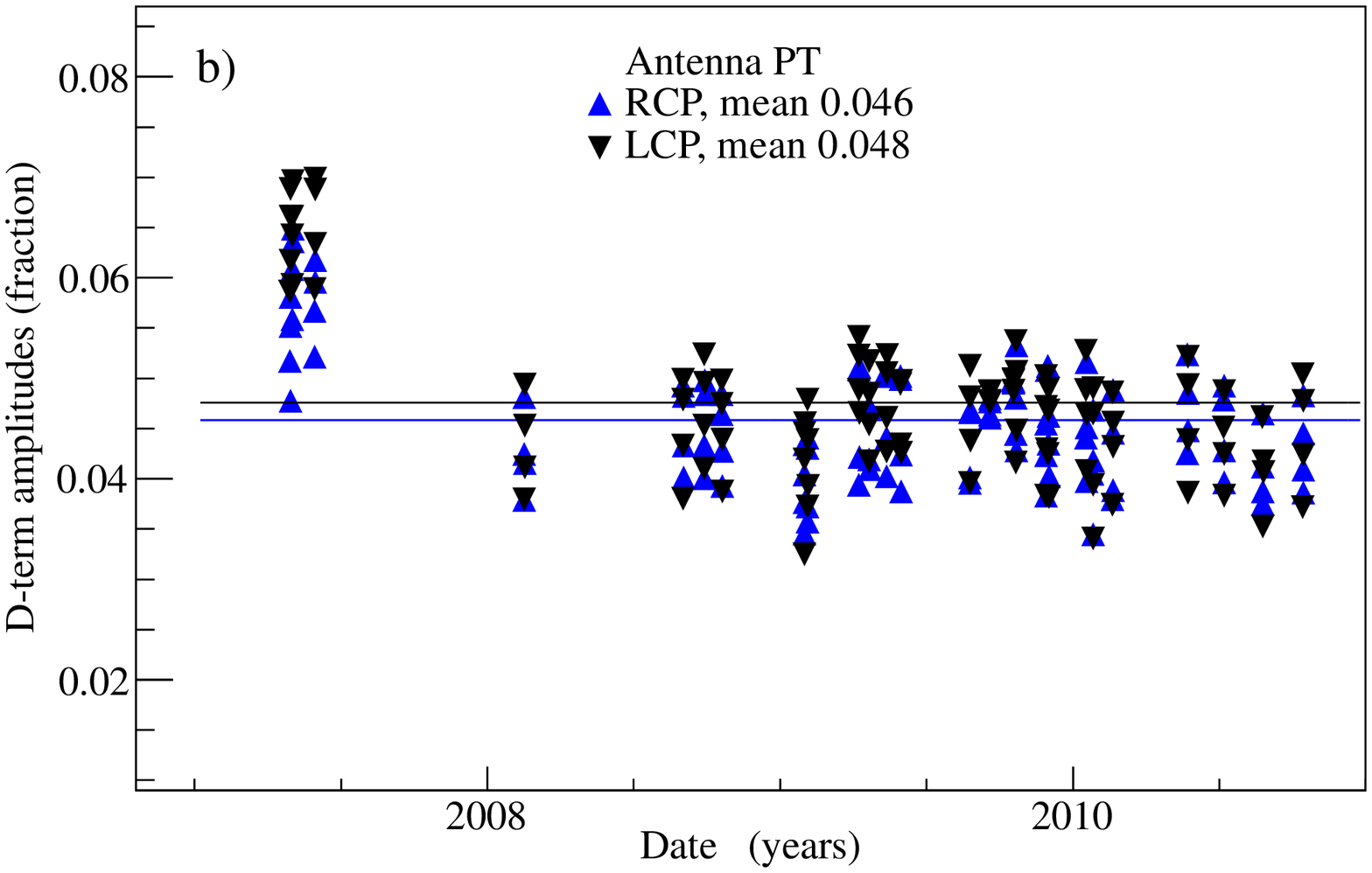}\\
\caption{(\textbf{a}) D-term right-left circular polarization (RCP) (blue triangles) and LCP
 (black inverted triangles) phases (\textbf{up}) and amplitudes (\textbf{bottom}) for PT
  antenna at 4.6~GHz, two IFs.
   Solutions for phases are adjusted to lie in a range $\pm180^{\circ}$. The mean values are given. (\textbf{b}) The same as in (\textbf{a}), but for 15.4~GHz.}
\label{f:1}
\end{figure}

An example of R and L D-term phases and amplitudes for the PT antenna at 4.6~GHz (two IFs) and 15.4~GHz (four IFs) is shown in Figure~\ref{f:1}.
The results for other antennas and frequencies are available upon request from E.K.
As can be seen, leakage solutions stay stable for almost four years.
To calibrate the single-epoch R-L phase difference we minimize the difference between D-term phases over all antennas and IFs and the average antenna- and IF-based solutions, determined by connecting all available epochs.
The~contribution of antennas with low SNR can be neglected, while they show larger variations in the D-term phases.
The~resulting accuracy in determining absolute EVPA is of 4$^{\circ}$, 3$^{\circ}$, 5$^{\circ}$ and 7$^{\circ}$ for 4.6--8.4~GHz, 15.4~GHz, 23.8~GHz and 43.4~GHz, respectively.
These values are comparable to those reported by, e.g., \citep{2005AJ....130.1389L,2012AJ....144..105H}, obtained using the same method.
The~values of uncertainties in our estimates mainly depend on sparse observations of calibrators, which we used to anchor absolute EVPA.
Meanwhile, this technique is still very effective.
Unfortunately, it is not applicable to observations made after upgrade of VLBA receivers.

\section{Quasar 0850$+$581}
\label{s:2}
\vspace{-6pt}
\subsection{Jet Parameters and Kinematics}
\label{s:2.1}

\begin{figure}[H]
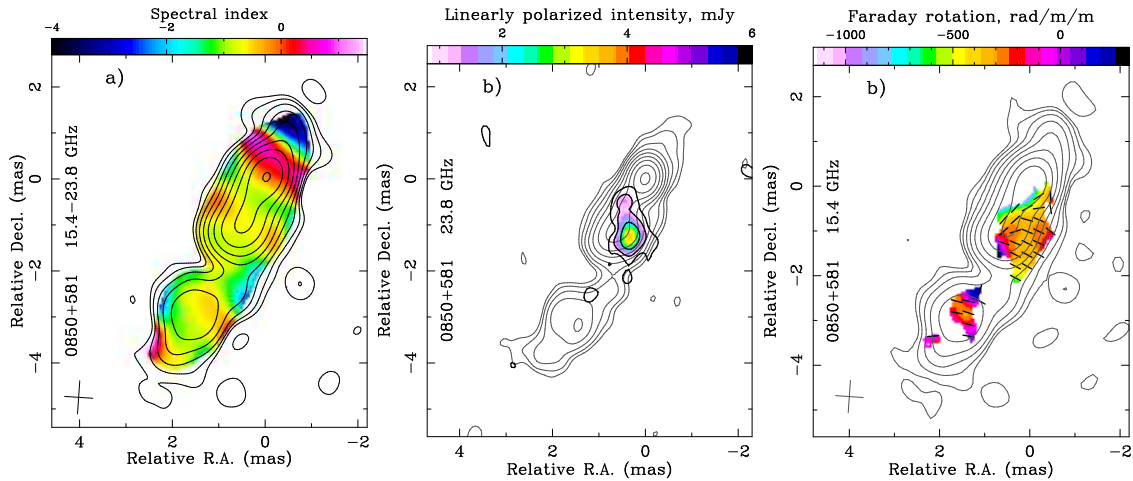

\centering
\includegraphics[width=6.1 cm,angle=270]{fig3}
\includegraphics[width=6.25 cm,angle=270]{fig4}
\includegraphics[width=6.25 cm,angle=270]{fig51}
\caption{(\textbf{a}) 0850$+$581 15.4--23.8~GHz spectral index distribution (color) overlaid on Stokes I levels (solid grey lines). Contours are drawn at ($-$1,1,2,4...)$\times$0.6~mJy/beam. Spectral index $\alpha$ is defined as $S\propto\nu^{\alpha}$, where $S$ is the flux density observed at frequency $\nu$.
(\textbf{b}) Contours of linearly polarized intensity (black solid lines) overlayed on Stokes I levels (grey solid lines) at 23.8~GHz and drawn at ($-$1,1,2,4...)$\times$0.46~mJy/beam.
(\textbf{c})~Faraday rotation measure, determined in the range 8.1--23.8~GHz, together with Stokes I levels (grey solid lines) and Faraday-corrected EVPA (black solid sticks) at 15.4~GHz, drawn at ($-$1,1,2,4...)$\times$0.6~mJy/beam.}
\label{f:4}
\end{figure}

\begin{figure}[H]
\centering
\includegraphics[width=8.3 cm,angle=0]{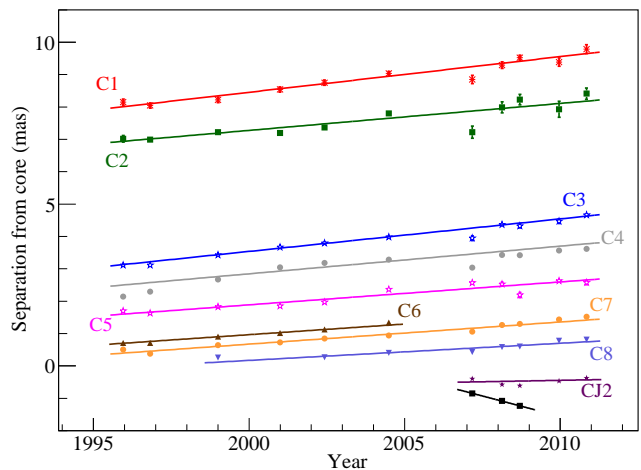}
\caption{The separation of the modelled jet components from the core with time at 15.4~GHz. The solid lines represent a linear fit to the data points. The nomenclature for jet components has been adapted from \cite{2016AJ....152...12L} and is shown by letters.}
\label{f:3}
\end{figure}

The twin-sided jet structure of the quasar has been revealed on VLBA Stokes~I contours at 15.4--43.3~GHz.
Meanwhile, the counter-jet is not visible at 4.6--8.4~GHz.
The spectral index distribution in the core region (fee Figure~\ref{f:4}) shows an inverted spectrum as expected for synchrotron self-absorbed emission.
The emission of the counter-jet is optically thin and significantly steeper than in the approaching jet.
The estimated ratio $R$ of flux densities for the jet and counter-jet accounts for at least $5.0\pm0.5$.
\vspace{-6pt}
The VLBA-MOJAVE program at 15.4~GHz delivers observations of the target at 10 epochs from 2004 to 2010.
This data has been combined with our observations to analyse quasar kinematics.
The~structure of the source is modelled by circular two-dimensional Gaussians on the visibility plane.
The~results (given in Figure~\ref{f:3}) show a linear increase in the velocity of the components with separation form the apparent jet base.
Starting from $(3.38\pm0.04)c$ at 0.3~mas from the core, the apparent velocity reaches $(7.0\pm0.4)c$ 9~mas downstream.
The preliminary results of the counter-jet structure show the presence of the two components, visible at three (CJ1) and five (CJ2) epochs of observations.
Both of them are detected at 23.8~GHz, while CJ1 is two faint and invisible at 43.3~GHz.

\subsection{Linear Polarization}
Linear polarization is not detected in the core and counter-jet, while it is visible in the approaching jet at all observed frequencies.
The polarized intensity is displaced from the jet axis towards its approaching edge, as shown in Figure~\ref{f:4} for 15.4~GHz.
The maximum degree of polarization accounts is (32 $ \pm$ 3)\% and is observed at the jet edge at 23.8~GHz.
It points to the presence of a well ordered magnetic field in the quasar jet.
Moreover, a closer look at this region at 43.4~GHz shows that the linear polarization that was detected there coincides with the point where Stokes~I contours have distorted shape.
While the jet significantly bends downstream, as discussed in Section~\ref{s:2.1}, it is likely that this polarization picture is produced by interaction of the jet with its surrounding medium.

\subsection{Faraday Rotation Measure}

To perform Faraday rotation measure (RM) imaging we convolved the images at lower frequencies with the synthesized beam matched to the intermediate band of the used range of wavelengths.
Comparing this to the conventional method, which uses the beam of lowest frequency, this approach minimizes smearing of the polarized intensity in a large beam at high frequencies, thus allowing the more precise location of the RM in the jet.
We compare the results of these approaches on different sources and in different wavelength ranges.
The analysis does not show discrepancy in the RM values between them, meanwhile the polarized contours that tapered with the intermediate beam are better reproduced in polarized intensity maps and convolved with their natural CLEAN beams.

The 0850$+$581 Faraday rotation measure was calculated over 4.6--8.4 and 8.1--23.8~GHz bands separately, and for higher range is shown in Figure~\ref{f:4}.
EVPA follows linear dependence with the squared wavelength well. Meanwhile, fractional polarization increases towards shorter wavelengths.
While~we estimate the Faraday rotation $\sim0.5$~mas downstream the core region of $\lesssim-4500$~rad/m$^2$ in the rest frame of the source, it reaches $-1612\pm102$~rad/m$^2$ at the point with the highest degree of polarization (2~mas from the core), changes sign and increases to $270\pm70$~rad/m$^2$ downstream (3.5~mas from the core).
Linear EVPA vs. $\lambda$ behavior and change in EVPA of $\sim60^{\circ}$ due to Faraday rotation at 20~GHz imply an external nature of magnetized medium relative to synchrotron emitting~regions.

\section{Discussion}

The minimum component speed $\beta_\mathrm{min}$ and maximum viewing angle $\theta_\mathrm{max}$ consistent with the estimated apparent velocity $\beta_\mathrm{app}$ are given by
\begin{equation}
\beta_\mathrm{min} \geq \frac{\beta_\mathrm{app}}{\sqrt{1+\beta_\mathrm{app}^2}}, \theta_\mathrm{max}=2\mathrm{atan}\frac{1}{\beta_\mathrm{app}}.
\end{equation}

Substituting extreme observed speeds of 3.3c and 7c, this results in $\beta_\mathrm{min,3.3c}\geq 0.96$, $\theta_\mathrm{max,3.3c}\leq 34^{\circ}$ and $\beta_\mathrm{min,7c}\geq 0.99$, $\theta_\mathrm{max,7c}\leq 16^{\circ}$, accordingly.
The~minimum jet to counter-jet flux density ratio can be estimated as
\begin{equation}
R = \big(\frac{1+\mathrm{cos}\theta_\mathrm{max}}{1-\mathrm{cos}\theta_\mathrm{max}}\big)^{2-\alpha},
\end{equation}
where $\beta=1$. It results in $R_{3.3c}\geq119$ and $R_{7c}\geq2400$, considering zero spectral index.
These estimates imply that the observed jet-to-counter-jet ratio of 5 is inconsistent with any of the observed apparent velocities.
There are two plausible scenarios which might explain this: (a) the ejection of components in the vicinity of the nucleus occurs with lower velocities, while acceleration takes place further downstream and (b)~the jet and counter-jet are seen at different viewing angles.

We registered high values of rotation measure from the regions, located 0.5--2~mas away from the core, which in deprojection corresponds to 12--50~pc.
This suggests the presence of dense ionized gas surrounding the nucleus of 0850$+$581.
The large inclination angle of the jet makes it possible to observe this medium, as the synchrotron emission passes through it.
This might be the disk wind, broad or narrow line regions~\citep{1989agna.book.....O}.
Linear polarization profiles indicate that this medium is located close to the jet emitting regions, since the excess is seen at a location where the jet bends and interacts with the surrounding medium.
We do not see significant changes in spectral index at this region, which implies that there are no opacity changes.
This dense gas may be responsible for the free-free absorption, which affects the emission in the counter-jet at low frequencies.
Meanwhile, the spectral index value at 5.0--15.4~GHz is $2$ which is consistent with both free-free and synchrotron self absorption.
The further analysis of brightness temperature distribution might make it possible to distinguish between them.

We will continue working with this quasar, as it has already shown a variety of unusual properties.
The full~results of this study will be published elsewhere.

\vspace{6pt}

\acknowledgments{
This work was supported by Russian Science Foundation (project 16-12-10481).
The Long Baseline Observatory is a facility of the National Science Foundation operated under cooperative agreement by Associated Universities, Inc.
This research has made use of data (absolute electric vector position angle calibration) from the University of Michigan Radio Astronomy Observatory which has been supported by the University of Michigan and by a series of grants from the National Science Foundation, most recently AST-0607523.
This study makes use of 43~GHz VLBA data from the VLBA-BU Blazar Monitoring Program, funded by NASA through the Fermi Guest Investigator Program.
This research has made use of data from the MOJAVE database that is maintained by the MOJAVE team \citep{lister_etal09}.
The luminosity distance of the source was computed using the cosmological calculator \citep{wright_06}.
}

\authorcontributions{Y.Y.K.\ has conceived and designed the experiment as well as performed initial calibration; E.K. has done phase-referencing and polarization calibration, analyzed the data; both authors have discussed results while E.K.\ has lead and written the paper.}

\conflictsofinterest{The authors declare no conflict of interest. The founding sponsors had no role in the design of the study; in the collection, analyses, or interpretation of data; in the writing of the manuscript, and in the decision to publish the results.}

\reftitle{References}


\begin{thebibliography}{999}
\providecommand{\natexlab}[1]{#1}

\bibitem[{Meier} \em{et~al.}(2001){Meier}, {Koide}, and
  {Uchida}]{2001Sci...291...84M}
{Meier}, D.L.; {Koide}, S.; {Uchida}, Y.
\newblock {Magnetohydrodynamic Production of Relativistic Jets}.
\newblock {\em Science} {\bf 2001}, {\em 291},~84--92.

\bibitem[{Roberts} \em{et~al.}(1994){Roberts}, {Wardle}, and
  {Brown}]{roberts_etal94}
{Roberts}, D.H.; {Wardle}, J.F.C.; {Brown}, L.F.
\newblock {Linear polarization radio imaging at milliarcsecond resolution}.
\newblock {\em {Astrophys. J.}} {\bf 1994}, {\em 427},~718--744.

\bibitem[{Leppanen} \em{et~al.}(1995){Leppanen}, {Zensus}, and
  {Diamond}]{1995AJ....110.2479L}
{Leppanen}, K.J.; {Zensus}, J.A.; {Diamond}, P.J.
\newblock {Linear Polarization Imaging with Very Long Baseline Interferometry
  at High Frequencies}.
\newblock {\em {Astron. J.}} {\bf 1995}, {\em 110},~2479, doi:10.1086/117706.

\bibitem[{G{\'o}mez} \em{et~al.}(2002){G{\'o}mez}, {Marscher}, {Alberdi},
  {Jorstad}, and {Agudo}]{gomez_etal02}
{G{\'o}mez}, J.L.; {Marscher}, A.P.; {Alberdi}, A.; {Jorstad}, S.G.; {Agudo},
  I.
\newblock {Polarization Calibration of the VLBA Using the D-terms.}
\newblock {\em VLBA Sci. Memo} {\bf 2002}, {\em 30}, 1-16.

\bibitem[{Kovalev} \em{et~al.}(2008){Kovalev}, {Lobanov}, {Pushkarev}, and
  {Zensus}]{kovalev_etal08}
{Kovalev}, Y.Y.; {Lobanov}, A.P.; {Pushkarev}, A.B.; {Zensus}, J.A.
\newblock {Opacity in compact extragalactic radio sources and its effect on
  astrophysical and astrometric studies}.
\newblock {\em {Astron. Astrophys.}} {\bf 2008}, {\em 483},~759--768.
  \href{http://xxx.lanl.gov/abs/0802.2970}{{\normalfont [0802.2970]}}.


\bibitem[{Kravchenko} \em{et~al.}(2017){Kravchenko}, {Kovalev}, and
  {Sokolovsky}]{kravchenko_etal17}
{Kravchenko}, E.V.; {Kovalev}, Y.Y.; {Sokolovsky}, K.V.
\newblock {Parsec-scale Faraday rotation and polarization of 20 active galactic
  nuclei jets}.
\newblock {\em {Mon. Not.  RAS}} {\bf 2017}, {\em 467},~83--101.
  \href{http://xxx.lanl.gov/abs/1701.00271}{{\normalfont
  [arXiv:astro-ph.HE/1701.00271]}}.

\bibitem[{Lisakov} \em{et~al.}(2017){Lisakov}, {Kovalev}, {Savolainen},
  {Hovatta}, and {Kutkin}]{lisakov_etal17}
{Lisakov}, M.M.; {Kovalev}, Y.Y.; {Savolainen}, T.; {Hovatta}, T.; {Kutkin},
  A.M.
\newblock {A connection between {$\gamma$}-ray and parsec-scale radio flares in
  the blazar 3C 273}.
\newblock {\em {Mon. Not. RAS}} {\bf 2017}, {\em 468},~4478--4493.
   \href{http://xxx.lanl.gov/abs/1703.07976}{{\normalfont [1703.07976]}}.

\bibitem[{Kravchenko} \em{et~al.}(2016){Kravchenko}, {Kovalev}, {Hovatta}, and
  {Ramakrishnan}]{kravchenko_etal16}
{Kravchenko}, E.V.; {Kovalev}, Y.Y.; {Hovatta}, T.; {Ramakrishnan}, V.
\newblock {Multiwavelength observations of the {$\gamma$}-ray flaring quasar S4
  1030+61 in 2009-2014}.
\newblock {\em {Mon. Not. RAS}} {\bf 2016}, {\em 462},~2747--2761.
   \href{http://xxx.lanl.gov/abs/1607.05852}{{\normalfont
  [arXiv:astro-ph.HE/1607.05852]}}.

\bibitem[{Bridle} and {Greisen}(1994)]{bridle_greisen_94}
{Bridle}, A.H.; {Greisen}, E.W.
\newblock {The NRAO AIPS Project---A Summary.}
\newblock {\em AIPS Memo} {\bf 1994}, {\em 87}, 1-3.

\bibitem[{Taylor} and {Myers}(2000)]{taylor_myers_00}
{Taylor}, G.B.; {Myers}, S.T.
\newblock {Polarization Angle Calibration Using the VLA Monitoring Program.}
\newblock {\em VLBA Sci. Memo} {\bf 2000}, {\em 26}, 1-13.

\bibitem[{Lister} \em{et~al.}(2009){Lister}, {Aller}, {Aller}, {Cohen},
  {Homan}, {Kadler}, {Kellermann}, {Kovalev}, {Ros}, {Savolainen}, {Zensus},
  and {Vermeulen}]{lister_etal09}
{Lister}, M.L.; {Aller}, H.D.; {Aller}, M.F.; {Cohen}, M.H.; {Homan}, D.C.;
  {Kadler}, M.; {Kellermann}, K.I.; {Kovalev}, Y.Y.; {Ros}, E.; {Savolainen},
  T.; et al.
\newblock {MOJAVE: Monitoring of Jets in Active Galactic Nuclei with VLBA
  Experiments. V. Multi-Epoch VLBA Images}.
\newblock {\em {Astron. J.}} {\bf 2009}, {\em 137},~3718--3729.
  \href{http://xxx.lanl.gov/abs/0812.3947}{{\normalfont [0812.3947]}}.

\bibitem[{Lister} and {Homan}(2005)]{2005AJ....130.1389L}
{Lister}, M.L.; {Homan}, D.C.
\newblock {MOJAVE: Monitoring of Jets in Active Galactic Nuclei with VLBA
  Experiments. I. First-Epoch 15 GHz Linear Polarization Images}.
\newblock {\em Astron. J.} {\bf 2005}, {\em 130},~1389--1417.
  \href{http://xxx.lanl.gov/abs/astro-ph/0503152}{{\normalfont
  [astro-ph/0503152]}}.

\bibitem[{Hovatta} \em{et~al.}(2012){Hovatta}, {Lister}, {Aller}, {Aller},
  {Homan}, {Kovalev}, {Pushkarev}, and {Savolainen}]{2012AJ....144..105H}
{Hovatta}, T.; {Lister}, M.L.; {Aller}, M.F.; {Aller}, H.D.; {Homan}, D.C.;
  {Kovalev}, Y.Y.; {Pushkarev}, A.B.; {Savolainen}, T.
\newblock {MOJAVE: Monitoring of Jets in Active Galactic Nuclei with VLBA
  Experiments. VIII. Faraday Rotation in Parsec-scale AGN Jets}.
\newblock {\em Astron. J.} {\bf 2012}, {\em 144},~105, doi:10.1088/0004-6256/144/4/105.
  \href{http://xxx.lanl.gov/abs/1205.6746}{{\normalfont
  [arXiv:astro-ph.CO/1205.6746]}}.

\bibitem[{Lister} \em{et~al.}(2016){Lister}, {Aller}, {Aller}, {Homan},
  {Kellermann}, {Kovalev}, {Pushkarev}, {Richards}, {Ros}, and
  {Savolainen}]{2016AJ....152...12L}
{Lister}, M.L.; {Aller}, M.F.; {Aller}, H.D.; {Homan}, D.C.; {Kellermann},
  K.I.; {Kovalev}, Y.Y.; {Pushkarev}, A.B.; {Richards},~J.L.; {Ros}, E.;
  {Savolainen}, T.
\newblock {MOJAVE: XIII. Parsec-scale AGN Jet Kinematics Analysis Based on 19
  years of VLBA Observations at 15 GHz}.
\newblock {\em {Astron. J.}} {\bf 2016}, {\em 152},~12, doi:10.3847/0004-6256/152/1/12.
  \href{http://xxx.lanl.gov/abs/1603.03882}{{\normalfont [1603.03882]}}.

\bibitem[{Osterbrock}(1989)]{1989agna.book.....O}
{Osterbrock}, D.E.
\newblock {\em {Astrophysics of Gaseous Nebulae and Active Galactic Nuclei}};
   University Science Books: Mill Valley, CA, USA,  1989.

\bibitem[{Wright}(2006)]{wright_06}
{Wright}, E.L.
\newblock {A Cosmology Calculator for the World Wide Web}.
\newblock {\em {Publ. ASP}} {\bf 2006}, {\em 118},~1711--1715.
  \href{http://xxx.lanl.gov/abs/astro-ph/0609593}{{\normalfont
  [astro-ph/0609593]}}.

\end{thebibliography}
\end{document}